\documentstyle[prb,aps,tighten,eqsecnum,floats,epsf]{revtex}
\preprint{T98/120} 
\begin{document} 
\draft 
\title{INCOMMENSURABILITY IN THE MAGNETIC EXCITATIONS
OF THE BILINEAR-BIQUADRATIC SPIN-1 CHAIN}
\author{O. Golinelli$^1$, Th. Jolic\oe ur$^1$
and E. S. S\o rensen$^{2}$\thanks{e-mail: 
golinel, thierry@spht.saclay.cea.fr,
sorensen@irsamc2.ups-tlse.fr}}
\address{$^{1}$ Service de Physique Th\'eorique, 
CEA Saclay,\\ F91191 Gif-sur-Yvette, France}
\address{$^{2}$ Laboratoire de Physique Quantique
IRSAMC, Universit\'e Paul Sabatier, \\ F31062 Toulouse, France}
\date{December 9, 1998} 
\maketitle
\begin{abstract} 
We study the magnetic excitation spectrum of the
spin-1 chain with Hamiltonian
${\cal H}=
\sum_{i} \cos \theta {\bf S}_{i}\cdot {\bf S}_{i+1}
+ \sin \theta ({\bf S}_{i}\cdot {\bf S}_{i+1})^{2}$.
We focus on the range $0\leq \theta \leq +\pi/4$
where the spin chain is in the gapped Haldane phase. 
The excitation spectrum and static structure factor is
studied using direct Lanczos diagonalization of small systems
and density-matrix renormalization group techniques
combined with the single-mode approximation.
The magnon dispersion has a minimum
at $k=\pi$ until a critical value $\theta_{c}=0.38$ is reached
at which the curvature (velocity) vanishes.
Beyond this point, which is distinct from the VBS point
and the Lifshitz point, the minimum lies at an incommensurate
value that goes smoothly to $k=2\pi/3$ when $\theta$ approaches
$\pi/4$, the Lai-Sutherland point. The mode
remains isolated from the other states~:
there is no evidence of spinon
deconfinement before the point $\theta =+\pi/4$.
These findings explain recent observation of the behavior
of the magnetization curve $M\approx (H -H_{c})^{1/4}$ 
for $\theta =\theta_{c}$. 
\end{abstract}
\pacs{\rm PACS: 75.10.Jm, 75.20.Hr} 

\section{Introduction}

It is now well known that one-dimensional spin-$S$ Heisenberg
antiferromagnets (AF) have qualitatively different properties
according to whether the spin value $S$ is integer or 
half-integer\cite{FDM}. The existence of a singlet-triplet gap
just above the ground state is clearly 
established\cite{white1,erik1,GJL,Uli} by numerical techniques
for the S=1 and S=2 cases. Our physical understanding of these
phenomena is based on the construction due to Affleck, Kennedy,
Lieb and Tasaki\cite{aklt} (AKLT). These authors were able to obtain 
explicitly
the ground state of the following bilinear-biquadratic 
Hamiltonian~:
\begin{equation}
{\mathcal H}_{aklt}= 
\sum_{i} {\bf S}_{i}\cdot {\bf S}_{i+1}
+ {1\over 3} ({\bf S}_{i}\cdot {\bf S}_{i+1})^{2},
\label{Haklt}
\end{equation}
where ${\bf S}_{i}$ are quantum $S=1$ spin operators
living on a chain whose sites are indexed by $i$.
To construct the ground state wavefunction of the Hamiltonian (1.1),
each original $S=1$ spin is written as two $S$=1/2 spins in a 
triplet state. Then the ground state is obtained by coupling
into a singlet state all nearest-neighbor spins-1/2,
thus forming a crystalline pattern of valence bonds. This
state is called the valence-bond-solid state (VBS).
It is in fact the unique ground-state and excitations
have a gap which is known rigorously to be nonzero.
Many results followed from the VBS picture. For example,
it implies that there are free spins 1/2 at the end of an open 
chain\cite{Kennedy}.
This has been verified experimentally\cite{jp,meisels} by electron 
spin resonance
on copper spins randomly introduced in the S=1 AF chain compound NENP.
The bulk excitations are easily pictured~: by breaking
a singlet bond into a triplet one creates a local objet that move along
the chain. This is only an approximate eigenstate but 
it has good
overlap\cite{knabe,AAH,FS,scharf} with the first excited state which 
is a triplet, the ``magnon'', as predicted
in the original derivation of Haldane\cite{FDM}. 

This appealing picture is certainly correct for the Hamiltonian 
(\ref{Haklt})
but it remains to be understood how closely it applies to the standard
Heisenberg exchange Hamiltonian for which the
biquadratic coupling in Eq.(\ref{Haklt}) is zero. Initial
studies\cite{script} proposed the generalized family of models~:
\begin{equation}
{\mathcal H}_{\theta}= 
\sum_{i} \cos \theta\, {\bf S}_{i}\cdot {\bf S}_{i+1}
+ \sin \theta \, ({\bf S}_{i}\cdot {\bf S}_{i+1})^{2}.
\label{Htheta}
\end{equation}
This family includes the familiar Heisenberg model for $\theta =0$
and the AKLT Hamiltonian for $ \tan \theta_{VBS} =1/3$. Since 
the gap does not go to zero in the interval $[0, \theta_{VBS} ]$
it is likely that there is no phase transition and
that the two limiting Hamiltonians $\theta =0$ and 
$\theta =\theta_{VBS}\simeq 0.3218 $ share the same physics.
It is exactly the same line of arguments that justify the use
of Laughlin wavefunctions to describe correlated states of electrons
in the Fractional Quantum Hall Effect\cite{Hall}.

However, the AKLT point has special 
properties~: for example the spin correlation
functions have a spatial decay which is purely exponential,
contrary to the Ornstein-Zernicke decay which is observed
for the Heisenberg model. A study of the family of models 
(\ref{Htheta}) revealed that the AKLT point is a {\it disorder}
point\cite{steph} beyond which short-range incommensurability 
appears in real space
spin correlations\cite{nous1}. Hence, it is at a boundary 
in some sense. The evolution of the spin excitations as a function
of $\theta$ is the subject of the present paper. For $\theta = \pi /4$
it is known from the Bethe Ansatz\cite{uimin,LS} that the spectrum
is gapless and that excitations form a continuum as in the case
of the spin-1/2 AF chain. Near the Heisenberg point, it is known
that there is an isolated branch of triplet excitations\cite{GJL2}
that enter in a two-particle continuum for a wavevector
$q\approx 0.3\pi$. In the neighborhood of $q=\pi$ the continuum
is well above the isolated mode. This holds also for the AKLT 
Hamiltonian, as shown by Lanczos diagonalization\cite{FS1}.
Since there is no phase transition with zero gap between
$\theta =0$ and the Lai-Sutherland point $\theta =+\pi/4$, 
the common belief
is that things evolve smoothly. This rather vague statement 
deserves however more scrutiny in view of the change of the
spin correlations at the AKLT point\cite{nous1}. An interesting issue
is the fate of the Haldane triplet mode that should ultimately
disappear in the spinon continuum for $\theta =\pi/4$. It has been
speculated recently\cite{yamamoto} that beyond the AKLT point
$\theta > \theta_{VBS}$ the Haldane mode disappears and is replaced
instead by  a gapped spinon continuum which becomes gapless
only right at the Lai-Sutherland point. However this intriguing 
picture is based only on variational wavefunctions whose relevance to 
the problem is at least unclear. It has also been 
observed\cite{okunishi} that the magnetization curve $M(H)$
displays intriguing behavior for some value of $\theta$. 
Some recent work\cite{GMu} has undertaken the study of dynamical 
properties between the Heisenberg and VBS points.

In this paper, we investigate the excitation spectrum for a range of
values of $\theta$ in the Haldane phase. Our main finding is that
the Haldane mode remains well-defined and isolated from other states~:
there is no deconfinement of spinons. The minimum wavevector
of the dispersion relation $E(k)$ stays at the zone boundary
$k=\pi$ till a critical value $\theta_{c}\approx 0.38$ 
which is {\it between} the
$\theta_{VBS}$ and Lifshitz points\cite{lifshitz} 
$\theta_{L}$ and clearly
different from these two values. Beyond $\theta_{c}$ the dispersion
has a minimum at some incommensurate wavevector that evolves smoothly
towards $2\pi /3$ when $\theta \rightarrow \pi/4$.
At $\theta_{c}$, the dispersion curve has a quartic minimum
and we show that this implies a
critical behavior $\approx (H-H_{c})^{1/4}$ for 
the magnetization $M(H)$.

In sect.~(\ref{lan}), we describe our results for the dispersion of 
the Haldane triplet obtained from Lanczos exact diagonalizations. 
In sect.~(\ref{SMA}), the single-mode approximation is applied
to DMRG calculations of the static structure factor in order
to estimate the lower edge of the excitation spectrum. 
We discuss the physical properties of the Haldane phase
and the relationship with the nonlinear sigma model
in sect.~(\ref{STR}). We discuss also the shape of the magnetization curve
in sect.~(\ref{STR}).
Finally
our conclusions are presented in sect.~(\ref{concl})

\section{Lanczos results}
\label{lan}
\begin{figure}
\epsfxsize=10cm	$$ \epsfbox{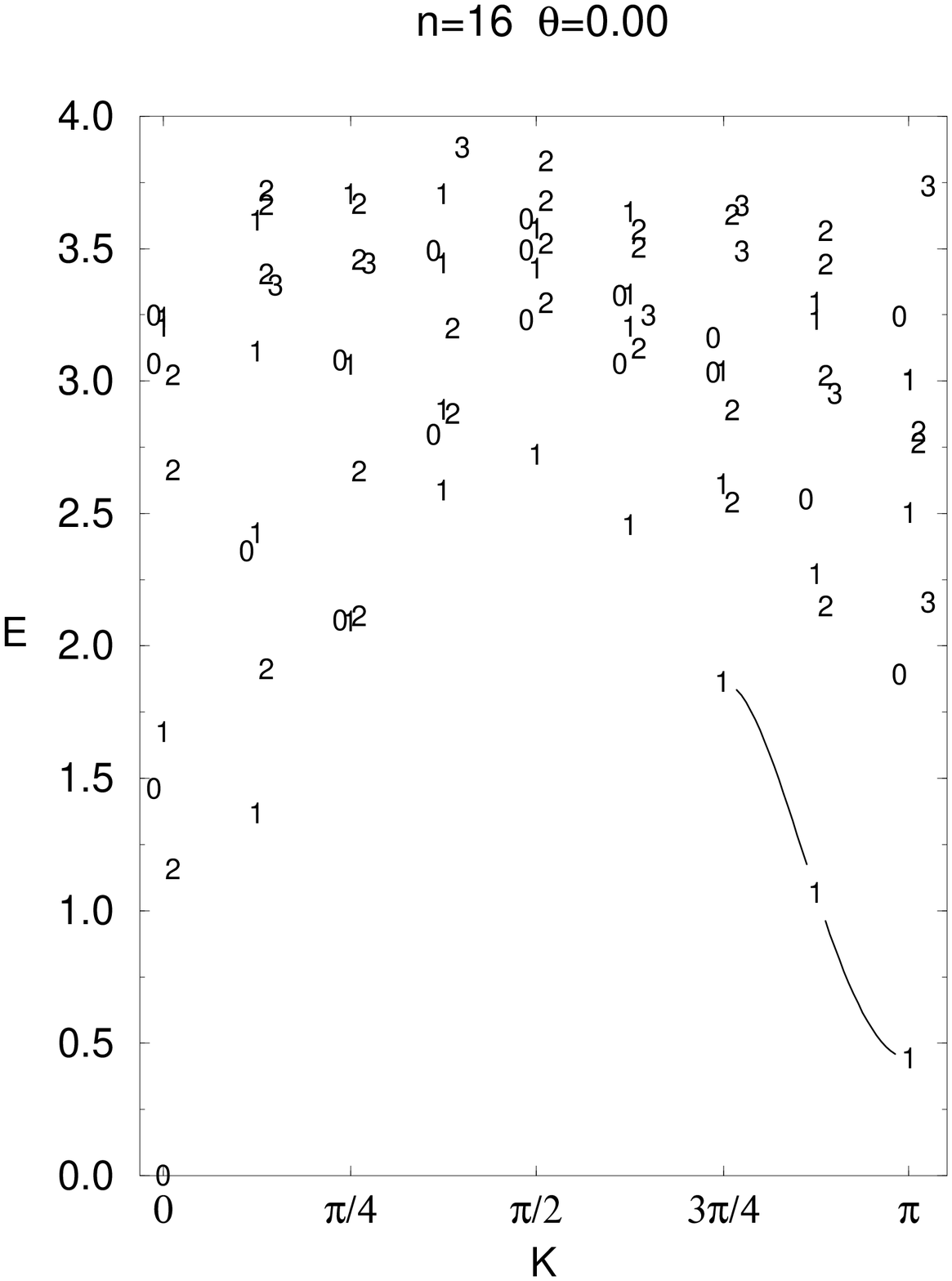} $$
\caption{Excitation spectrum of the N=16 chain for 
the Heisenberg S=1 AF chain, $\theta=0$.
Energies are relative to the ground state.
The number denote the total spin value of each state.
We have introduced small horizontal shifts to avoid
superposition of symbols.
The energies are given as a function of the chain momentum $k$
(only the first 10 levels are calculated for each momentum and
$S^{z}$).
Note the prominent triplet branch that starts from the
Haldane gap at the zone boundary at $k=\pi$.
The solid curve is the fit by Eq.~(\ref{fit}).}
\label{fheis}
\end{figure}

At the Heisenberg point $\theta =0$, the excitation spectrum of
the hamiltonian (\ref{Htheta}) has been studied in considerable detail.
We first use the Lanczos diagonalization method to study the interval
$0\leq \theta <+\pi/4$. We use conservation of
the z-component of the total spin  and compute the eigenenergies
of the first 10 low-lying states. 
We use also translation symmetry to work
in sectors with fixed lattice momentum $k$. To keep the hamiltonian
real even when the momentum $k$ is not 0 or $\pi$, we use 
a non-trivial change of basis due to Takahashi\cite{taka}.
This requires up to 350
Lanczos iterations which is considerably more expensive than
just getting the ground state of each  $S^{z}$ sector.
To prevent runaway of the Lanczos process we are forced
to re-orthogonalize with respect to the previous basis vector 
at each step.
This bottleneck leads to a limit on the length of the chain
which is 16 sites.
This is a difficulty which does not appear when computing only
the ground state. 
Our results for $\theta =0$ are shown in Figure~\ref{fheis}. 
Above the ground 
state, there is a well-defined triplet branch which is the 
prominent Haldane mode, the ``magnon''. 
The minimum of the dispersion is 
at the zone boundary $k=\pi$. This is the Haldane gap.
The mode enters a continuum\cite{GJL2} at $k\approx 0.3\pi$.
At $k=0$ the continuum is the bottom of the 
spectrum and starts at twice the Haldane gap (approximately
for N=16 sites on our figure but checked with excellent 
precision). 

There is a nonzero curvature of the dispersion $E(k)$ at $k=\pi$.
We model the bottom of the dispersion relation 
by a relativistic formula 
as suggested by the nonlinear sigma model~:
\begin{equation}
	E(k)=\sqrt{\Delta^{2}+ v^{2}(k-k_{0})^{2}}.
	\label{Ek1}
\end{equation}
Here $k_{0}$ is wavevector of the minimum,
$\Delta$ is the gap and $v$ is the velocity. Close to the minimum
we have~:
\begin{equation}
	E(k)\approx \Delta + {v^{2}\over 2\Delta}(k-k_{0})^{2}.
	\label{Ek2}
\end{equation}
The second derivative of the dispersion is thus simply
related to the velocity that occurs in the nonlinear sigma model
description.

\begin{figure}
\epsfxsize=10cm	$$ \epsfbox{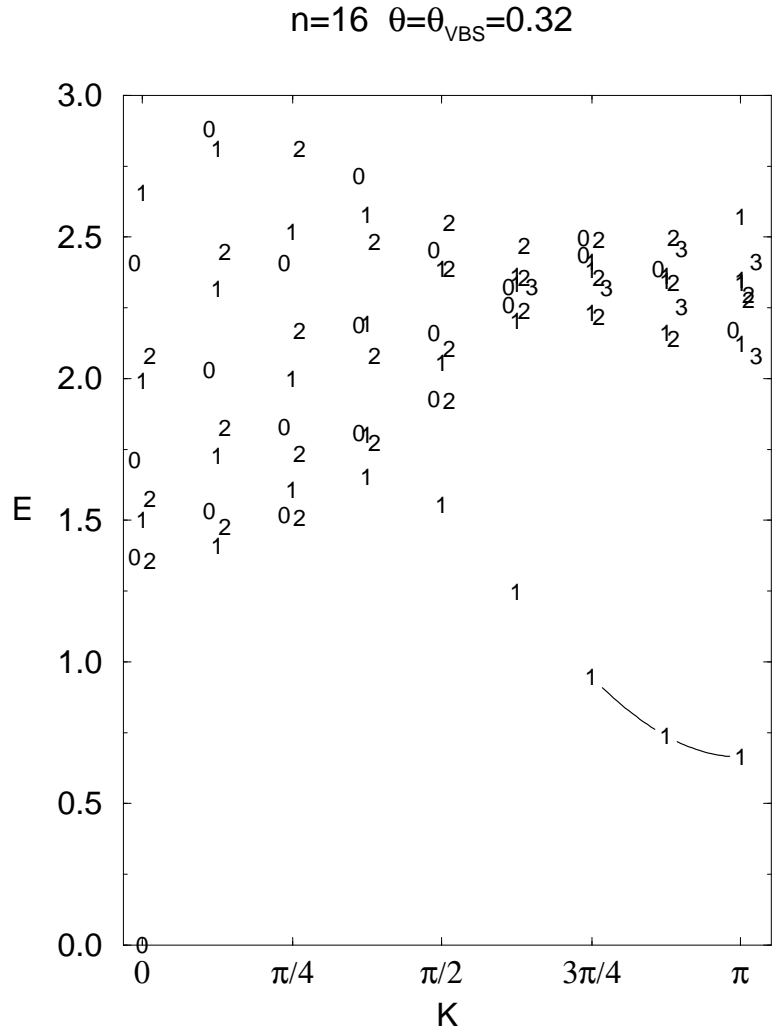} $$
\caption{Excitation spectrum of the N=16 chain for $\theta 
=\theta_{VBS}$. 
The physics is qualitatively similar to $\theta =0$, there is evidence
for a nonzero curvature of the dispersion at $k=\pi$
of the magnon and there is no spinon continuum.}
\label{fvbs}
\end{figure}

\begin{figure}							   
\epsfxsize=10cm	$$ \epsfbox{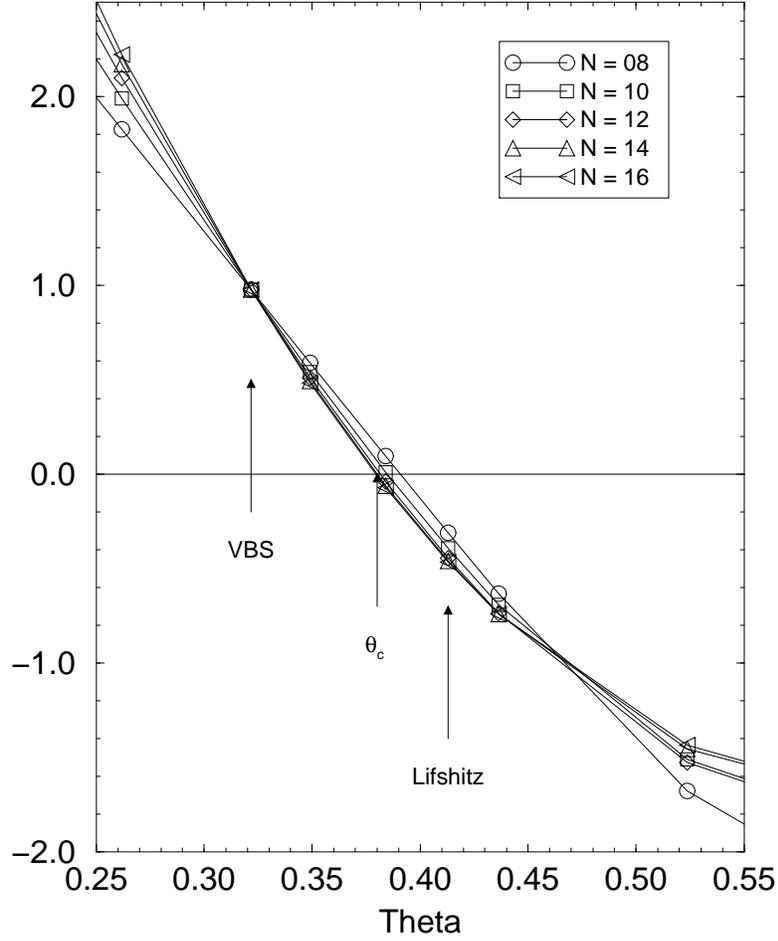} $$   
\caption{The second derivative $c$ of the dispersion relation
computed with the triplet state at $k=\pi$ and its two closest neighbors
along the magnon branch. Data are for several sizes N=8,10,12,14,16.
It is nonzero at the VBS point but vanishes for $\theta_{c}=0.38$.
Note that finite-size effects are extremely small in this range.
We find $c_{VBS}=0.9778(1)$.}
\label{veloce}							   
\end{figure}

When we increase $\theta$, there is a range of values for which
nothing qualitatively new happens. This range {\it includes}
the VBS point $\theta =\theta_{VBS}\simeq 0.3218$. The spectrum at 
this point is displayed in Figure~\ref{fvbs}. It shows the
same features as that of Figure~\ref{fheis}. 
Notably there is still a nonzero
curvature at $k=\pi$ of the magnon branch. 
If we consider the energy of the state at $k_{0}=\pi$ and
the two closest states on the magnon branch with $k=\pi - 2\pi/N$
and $k=\pi -4\pi/N$,
we can fit by a fourth-order polynomial~:
\begin{equation}
	E(k)=E(k_{0})+{c\over 2}(k-k_{0})^{2} +
	{d\over 24}(k-k_{0})^{4}.
	\label{fit}
\end{equation}
 and hence obtain an estimate of the 
velocity. Close to the VBS point, the values of $c$ as a function of
the chain length are given in Figure~\ref{veloce}.
At this point it is important to note 
that the finite-size effects are in fact extraordinarily small.
This is related to the fact that the spin-spin correlations in the VBS 
wavefunction are of very short range $\xi = 1/\log 3\approx 0.9$.
So our estimate for $c$ is very precise~: from N=4 to N=16 sites
the dependence is very weak. We can exclude a vanishing velocity
at this point. Indeed we find $c=0.9778(1)$.
Similarly the Haldane mode stays isolated from
the continuum which lies much above~: these states display
extremely fast convergence to the thermodynamic limit.

\begin{figure}
\epsfxsize=10cm	$$ \epsfbox{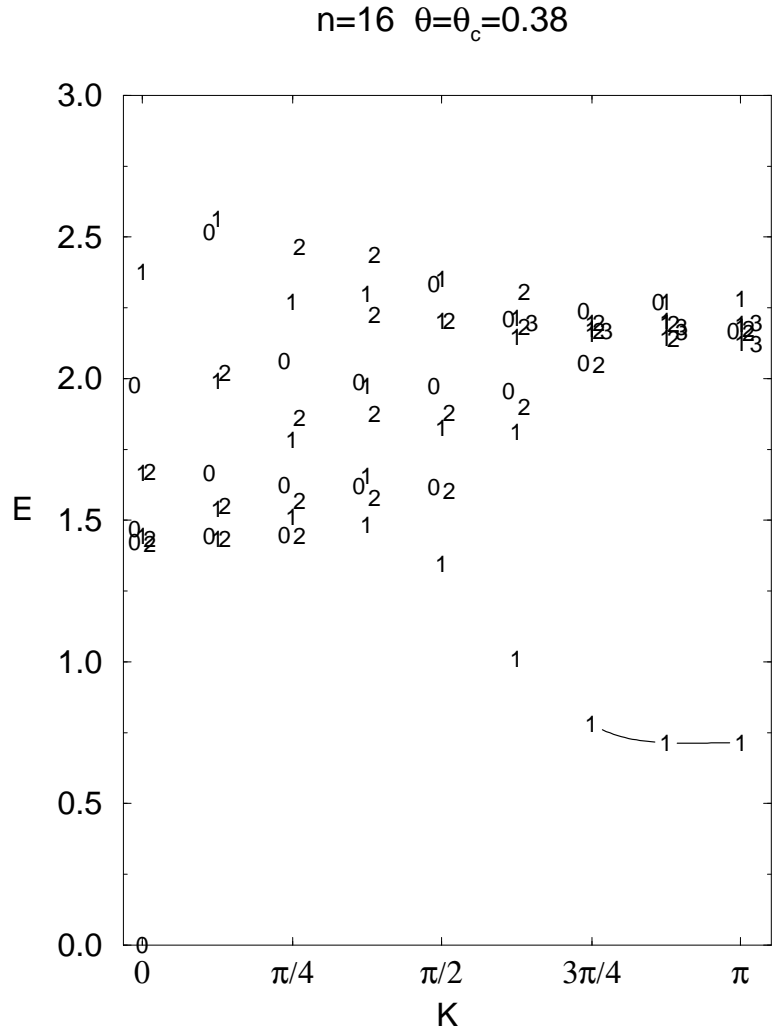} $$
\caption{Same spectrum as figure 1 but for $\theta =\theta_{c}$.
At this point the dispersion of the triplet mode has a fourth-order
minimum at the zone boundary.}
\label{fthetac}
\end{figure}

We observe in Figure~\ref{veloce} that the velocity vanishes at a point 
$\theta_{c}=0.38$
which is between the VBS point $\theta_{VBS}=0.32$
and the Lifshitz point 
$\theta_{L}=0.414$. There the
minimum of the dispersion relation becomes of fourth order
but still there are no states coming from above to fill the gap
between the Haldane mode and the continuum. 
This case is shown in Figure~\ref{fthetac}.
Once again we are in a regime of very weak finite-size effects~:
measurements of the spin correlation in this regime\cite{nous1}
are in agreement with the present findings.

\begin{figure}						   
\epsfxsize=10cm	$$ \epsfbox{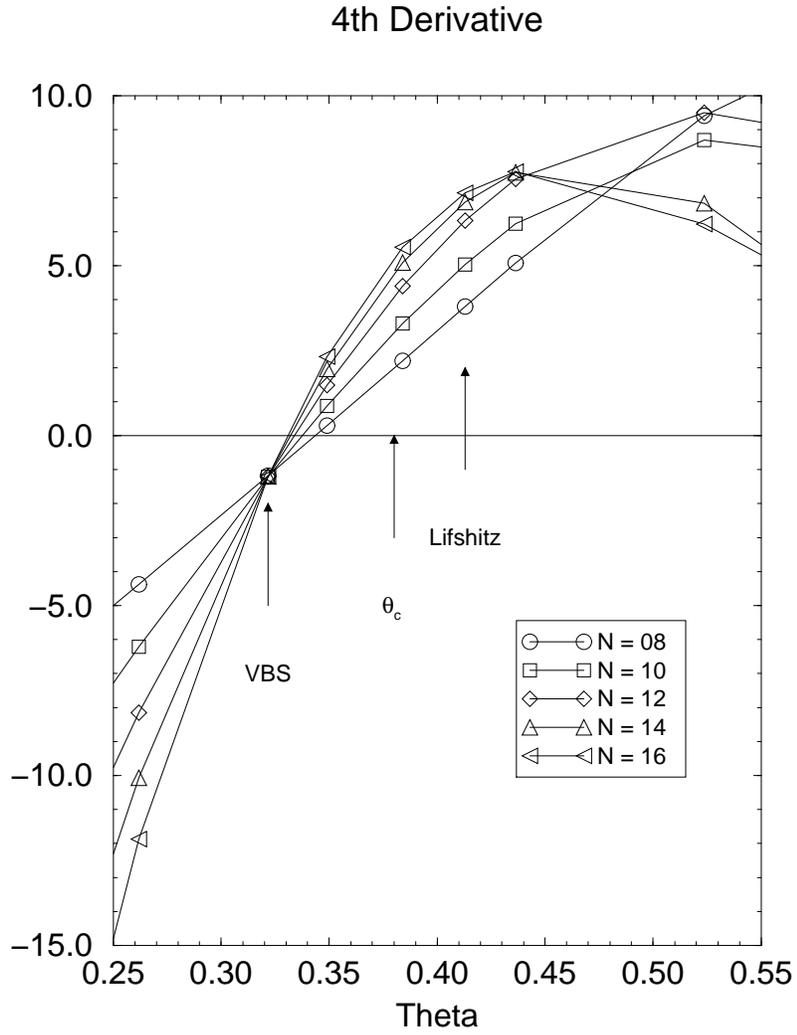} $$   
\caption{The fourth-order derivative $d$ of
the dispersion relation evaluated as in fig.(\ref{veloce})
with the two closest neighbors. It is nonzero at $\theta_{c}$.}	
\label{deriv4}					   
\end{figure}

We can also obtain an estimate of the 
fourth derivative $d$ of 
the dispersion relation $E(k)$ through Eq.~(\ref{fit}).
It is given in the neighborhood
of the VBS point in Figure~\ref{deriv4}. The size dependence is minimal 
again at $\theta_{VBS}$. Note that right at $\theta_{c}$ the fourth
derivative $d$ clearly extrapolate to a positive non-zero value~:
this fact will be important to explain the shape of the magnetization 
curve in sect.(\ref{STR}).

\begin{figure}
\epsfxsize=10cm	$$ \epsfbox{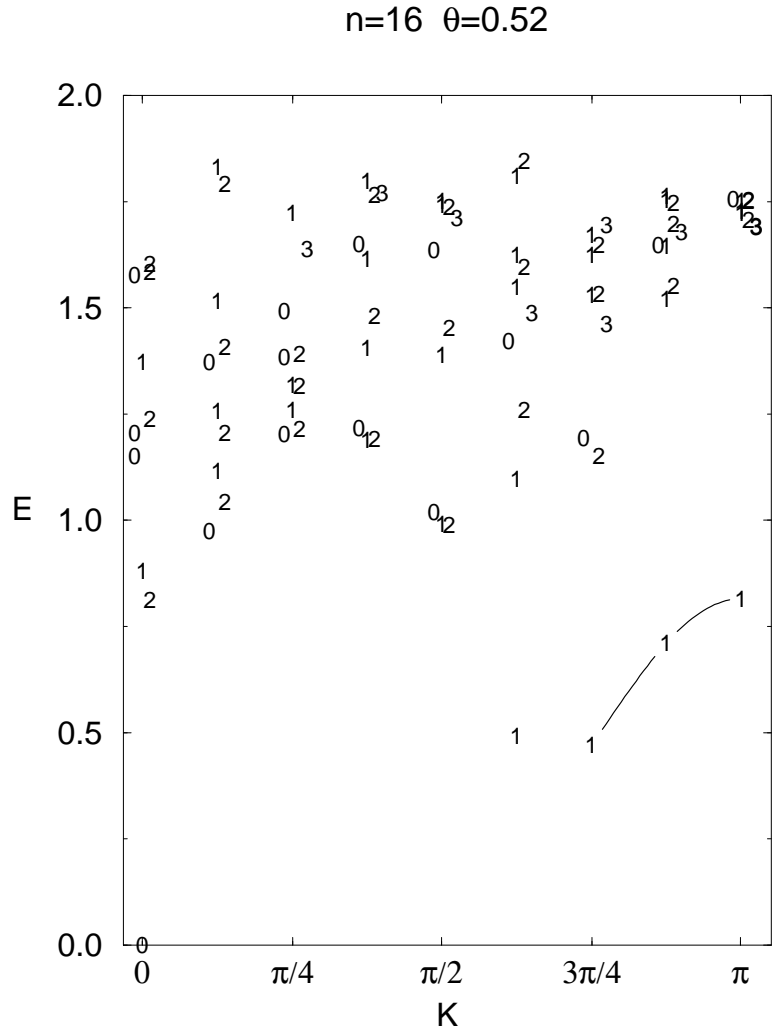} $$
\caption{Same spectrum as figure 1 for $\theta = 0.52$.
This point lies deep in the incommensurate regime. The minimum
wavevector is now inside the Brillouin zone and the magnon
mode remains isolated~: there is no evidence for spinon deconfinement.}
\label{finco}
\end{figure}

Beyond the special point $\theta_{c}$, the minimum of the 
dispersion no longer lies at $k=\pi$ but is at a wavevector
which is in general incommensurate and dependent upon $\theta$.
The curvature of the dispersion is now negative at the zone boundary~:
see Figure~\ref{veloce}.
A typical case is shown in Figure~\ref{finco} for $\theta = 0.52$.
There is no evidence for the deconfined spinons
proposed by Yamamoto\cite{yamamoto}. 
If this was the case, the corresponding energy
states should be filling the 
empty interval above the Haldane triplet mode.
We see no evidence for this.

When $\theta \rightarrow \pi/4$ the gap closes and the overall
picture is consistent with the apparition of the tripled period
$2\pi /3$. However in this regime finite-size effects become important
as the phase transition is approached.
It is known that right at
the Lai-Sutherland point, 
the phase transition is described by a 
a SU(3) generalization of the
Kosterlitz-Thouless phase transition\cite{Itoi} 
For $\theta > \pi/4$, the system enters a gapless phase.
In the regime $-\pi /4< \theta < \pi /4$, it is extremely likely that
the gap vanishes only at $\theta =\pm \pi/4$\cite{nous1,FS}.
However, we cannot exclude from our numerical study
that spinon deconfinement appears before that point
but this would involve yet another phase transition
for which there is at the present time no evidence.

\section{single-mode approximation}
\label{SMA}
To confirm the results obtained in the previous section,
we now turn to the single-mode approximation
of the dispersion relation. This approximation was introduced
originally by Bijl and Feynman to determine the phonon-roton
dispersion in superfluid $^{4}$He. It has been also successfully applied
to Haldane gap systems\cite{AAH}.

The dynamical structure factor of the spin system is defined by~:
\begin{equation}
	S(q, \omega )= \int^{+\infty}_{-\infty} dt
	e^{i\omega t} \sum_{n} e^{iqn} 
	\langle S^{\alpha}_{n}(t) S^{\alpha}_{0}\rangle .
	\label{Sdyn}
\end{equation}
In this equation, the $\alpha$ index need not be specified since
we consider only isotropic systems.
The equal-time correlation function is~:
\begin{equation}
	S(q)= \langle S^{\alpha}_{q} S^{\alpha}_{-q} \rangle
	= {1\over 2\pi} \int^{+\infty}_{-\infty} d\omega
	S(q, \omega ) ,
	\label{Stat}
\end{equation}
where $S^{\alpha}_{q} =1/\sqrt{L}\sum_{n} e^{iqn} S^{\alpha}_{n}$.
Imagine that we know the ground state $|0\rangle$. Then
a guess for the first excited state may be simply
$S^{\alpha}_{q}|0\rangle$. This is a triplet with lattice momentum $q$.
This will be close to an exact excited state if the dynamical
structure factor $S(q, \omega )$ is strongly peaked at the 
energy $\omega_{SMA}(q)$ of the state $S^{\alpha}_{q}|0\rangle$.
This energy is given by the formula~:
\begin{equation}
	\omega_{SMA}(q)= {1\over 2\, S(q)}
	\langle \left[S^{\alpha}_{-q},
	 \left[{\mathcal H_{\theta}},S^{\alpha}_{q}\right]\right]
	 \rangle ,
	\label{sma}
\end{equation}
where $\mathcal H_{\theta}$ is the Hamiltonian. The advantage of the 
single-mode approximation is that it does not require a full
dynamical study of the system~: in Eq.(\ref{sma}), it is
possible to evaluate all quantities once we have a good
approximation for the ground state. This is feasible 
by the DMRG algorithm which is extremely efficient in quantum spin 
chains.
The commutators in Eq.(\ref{sma}) can be evaluated
straightforwardly with the result~:
\begin{equation}
	\omega_{SMA}(q)=(\cos q -1)c(\theta )/S(q) ,
	\label{}
\end{equation}
where $c(\theta)$ is some constant given by~:
\begin{eqnarray}
		c^x(\theta) & = & \cos \theta 
		\left[ <S^y_iS^y_{i+1}>+<S^z_iS^z_{i+1}>\right]
        +\sin \theta [2<(\vec S_i\cdot \vec S_{i+1})^2>-
        <(\vec S_i\cdot \vec S_{i+1})S^x_iS^x_{i+1} 
        +S^x_iS^x_{i+1}(\vec S_i\cdot \vec S_{i+1})>
         \nonumber \\
        & &
        \mbox{}
        +2<S^y_iS^z_iS^z_{i+1}S^y_{i+1}+S^z_iS^y_iS^y_{i+1}S^z_{i+1}>
        -2<(S^z_i)^2(S^y_{i+1})^2+(S^y_i)^2(S^z_{i+1})^2>].
\end{eqnarray}
We find that $c(\theta)$ only varies slowly with $\theta$
and it is now obvious that at least within the 
single-mode approximation the minimum in
$\omega_{}(q)$ will start to move away from $\pi$ before the Lifshitz point
is reached (at the Lifshitz point the maximum of $S(q)$ moves away from 
$\pi$).
At the VBS point the ground state $|VBS\rangle$ is simple enough
that $\omega_{SMA}(k)$ can be calculated exactly\cite{AAH}.
One has $\omega_{SMA}(k)= {\sqrt{5}\over 9\sqrt{2}}(5+3\cos k)$.
Hence, we see explicitly that at the VBS point the second derivative
and hence the velocity is
non-zero within the SMA approximation. The value of the second
derivative $c_{SMA}=1.054$ is very close to our Lanczos
estimate of sect.~II $c=0.9778$.

\begin{figure}
\epsfxsize=10cm	$$ \epsfbox{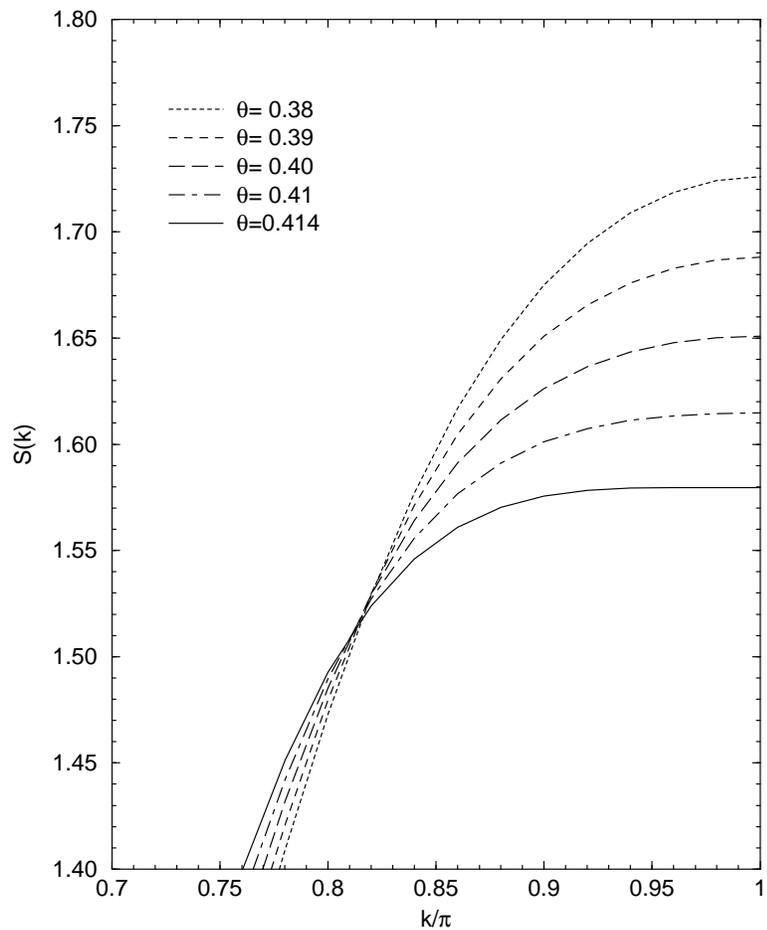} $$
\caption{The static structure factor from a DMRG calculation.
It is drawn only near the zone boundary for several $\theta$ values
in the interval $\theta_{c}, \theta_{L}$.}
\label{Sq}
\end{figure}

\begin{figure}
\epsfxsize=10cm	$$ \epsfbox{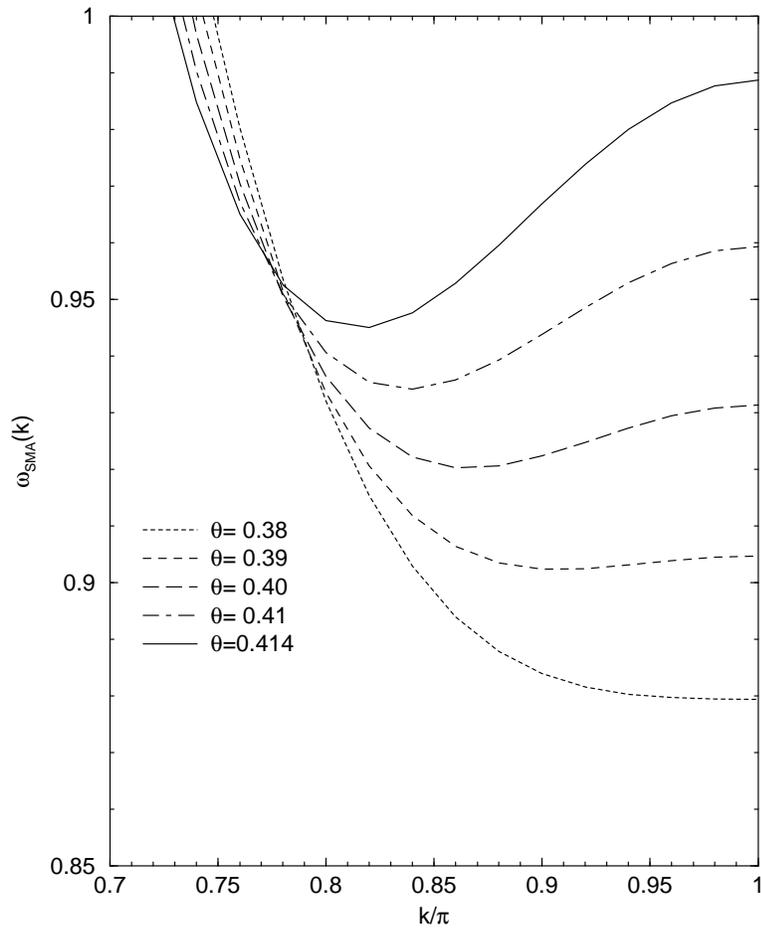} $$
\caption{The dispersion $\omega_{SMA}(k)$ obtained from the single-mode
approximation. The minimum moves away from $k=\pi$ at the value
$\theta_{c}$ which is the same as the Lanczos value $0.38(1)$.
Energies have been divided by $\cos\theta$.}
\label{ESMA}
\end{figure}

Using the above expressions, $c(\theta)$ as well as $S(q)$
have been calculated 
with the DMRG. We present results in the interval
between $\theta_{c}$ as defined in sect.(\ref{lan}) and $\theta_{L}$.
In Figure~\ref{Sq}, the structure factor is plotted in the neighborhood
of the zone boundary $q=\pi$. It becomes very flat, i.e. of fourth 
order at the Lifshitz point $\theta_{L}\simeq 0.414$.
However, the minimum in $\omega_{SMA}(q)$ 
starts to move
away from $\pi$ for $\theta\in[0.38, 0.39]$~: this can be seen on 
Figure~\ref{ESMA} where this quantity is plotted close to $q=\pi$.
This is consistent with our findings from
the exact diagonalization results. 
Clearly, $\theta_{c}$ and $\theta_{L}$ are distinct, and in fact
a considerable interval occurs between these two points.

Inherently the SMA approximation
assumes that
$$S(k, \omega) = S(k) \delta(\omega - E(k))$$
and hence neglects any incoherent background. Thus,
$\omega_{SMA}(k)$ {\it always} over estimates the true excitations.
This is important to remember when one compares the SMA results for
the gap with other numerical estimates. Roughly the distance that the SMA
value is above the true gap measures how big the incoherent contribution
to $S(k,\omega)$ is. It works quite well around the Lifshitz point.

\section{structures within the Haldane phase and magnetization process}
\label{STR}

From the numerical studies, both Lanczos and DMRG for single-mode 
approximation, we obtain a rather intricate structure within
the Haldane phase $-\pi/4<\theta < +\pi/4$.

$\bullet$ In the interval $-\pi /4 < \theta \leq \theta_{VBS}$,
the dispersion relation has its minimum at $k=\pi$,
the spin correlations are commensurate~: 
$\langle {\bf S}_{0}\cdot {\bf S}_{x} \rangle\simeq
(-)^{x}\exp (-x/\xi)/\sqrt{x}$.
In this regime the O(3) non-linear sigma model
is a perfectly correct description of the low-energy
long-wavelength behavior of the system in agreement with
the original derivation by Haldane\cite{FDM}.

$\bullet$ In the interval $\theta_{VBS} < \theta \leq \theta_{c}$,
the dispersion relation still has its minimum at $k=\pi$
but now the spin correlations oscillate with a period which
is incommensurate\cite{nous1}, This is not seen in the static
structure factor S(q) which remains peaked at $q=\pi$, a feature
due to the short-range nature of the spin order. While $E(k)$ is still
sigma model like, this regime is {\it not} described by the sigma model
which has only commensurate spin correlations. Indeed
the nonlinear sigma model uses the Ansatz~:
\begin{equation}
	{\bf S}_{i}=(-)^{i}{\bf \phi_{i}}+ {\bf L_{i}},
	\label{nlsm}
\end{equation}
where  the fields $\bf\phi$ and $\bf L$ are smooth i.e.
have Fourier modes only near $k=0$. This is no longer valid
even without a phase transition as a function of $\theta$.

$\bullet$ In the interval $\theta_{c} < \theta \leq \theta_{L}$,
the dispersion has now a minimum away from the zone boundary.
In fact there are two nonequivalent wavevectors $\pi\pm Q$ at the minimum.
The quantity S(q) is still peaked at $q=\pi$. A possible
effective theory would be now a non-linear sigma model
describing helical order\cite{Dombre,aza}. It would involve 
a 3 $\times$ 3 rotation matrix as an order parameter to describe
the short-range spin order. It is a phase with incommensurate
short-range spin correlations and the spinon are {\it confined}.

$\bullet$ In the interval $\theta_{L} < \theta \leq +\pi /4$,
the only difference with the previous case is that
there is a double-peak structure in S(q).

The above observations have interesting consequences for the 
magnetization curve M(H). For the Heisenberg model, its shape
and essential characteristics are well understood\cite{IBose,erik1}.
When the applied uniform magnetic field H is less than the Haldane gap,
there is no net magnetization~: since H is coupled to the z-component
of the total spin, the Zeeman Hamiltonian commutes with the Heisenberg
Hamiltonian and hence does not change wavefunctions. So the ground
state is unaffected by H. However at a critical value H$_{c1}$ = 
$\Delta$, the Haldane gap, there is a crossing of levels and 
a magnetized state becomes the ground state. Finally there is full
saturation to a ferromagnetic state beyond some field H$_{c2}$.
Near H$_{c1}$, the critical behavior is given by 
$M(H)\simeq (H-H_{c1})^{1/2}$. The exponent 1/2 is directly
related to the dispersion relation of the magnons near $k=\pi$.
Indeed the magnetization is given by inverting the relation 
$H-H_{c1}=dE/dM$ and the function E(M) is the same as for a system of
noninteracting fermions~:
\begin{equation}
	E = (\Delta -H) M +N \int^{+k_{F}}_{-k_{F}} {dk\over 2\pi}
	{v^{2}k^{2}\over 2\Delta}.
	\label{EM1}
\end{equation}
In this equation, N is the number of sites and
 $v$ is the velocity which is defined through
the dispersion relation~:
\begin{equation}
	E(k)=\Delta + {v^{2}\over 2\Delta}(k-\pi)^{2} + c_{4}(k-\pi)^{4}
	+O(k-\pi)^{6}.
	\label{V}
\end{equation}
Since $M=Nk_{F}/\pi$, one has~:
\begin{equation}
	E(M) =M(\Delta -H) + {(v\pi)^{2}\over 6\Delta}{M^{3}\over L^{2}}
	+d_{4}M^{5}+O(M^{7}).
	\label{EM2}
\end{equation}
Near the critical field this leads to $M\simeq (H-\Delta )^{1/2}$.
This depends crucially on the quadratic dispersion relation.
If now we consider the generalized Hamiltonian at $\theta=\theta_{c}$,
one has $v=0$ but the fourth-order derivative is nonzero as can be 
seen in Figure~\ref{deriv4}. Hence Eq.(\ref{V}) should be replaced
by $E(k)=\Delta + c_{4}(k-\pi)^{4}$. There is no cubic term
by imposing invariance under $k\rightarrow 2\pi -k$
which is parity and lattice periodicity. Now in Eq.(\ref{EM2}),
the $M^{4}$ term dominates and this leads immediately to
$M\simeq (H-\Delta )^{1/4}$. Our findings explain the observation
of such a behavior by Okunishi et al.\cite{okunishi}.
When $\theta >\theta_{c}$ the dispersion has again a {\it quadratic} 
minimum
and the line of arguments with $v\neq 0$ is again valid~:
$M\simeq (H-\Delta )^{1/2}$ is the correct behavior in this range.
The massless phase above $H_{c1}$ however is likely to be\cite{Gabor}
a two-component Luttinger liquid, contrary to the one-component
Luttinger liquid that arises for $\theta < \theta_{c}$.

\section{Conclusion}
\label{concl}
We have studied the magnetic excitation spectrum of the
bilinear-biquadratic S=1 Hamiltonian that includes
the Heisenberg point as well as the VBS point. We have shown
by Lanczos and DMRG techniques combined with the single-mode approximation
that the magnon dispersion becomes incommensurate at a critical value
$\theta_{c}=0.38$ which is different from all previously
known points in the phase diagram.
Right at the special value $\theta_{c}$ the magnon dispersion has a 
fourth-order minimum at the zone boundary $k=\pi$. As a consequence
the magnetization curve $M(H)$ near the 
lower critical field is drastically modified~: 
instead of a square-root behavior, the 
magnetization rises with a power-law 1/4. This explains recent 
observations.

From our study it is now clear that there are hidden structures
within the bulk of the Haldane phase. This phase extends in the 
interval $-\pi/4<\theta < +\pi/4$ and the gap vanishes at the two
boundary points. These critical points correspond, as is well known,
to transitions towards completely different phases. It is 
widely believed that absence of phase transition means that the physics
evolves smoothly, hence, that the Heisenberg point $\theta =0$
and the AKLT point $\theta =0.32$ share the same physical 
properties. It was known that spin correlations do change 
qualitatively even within the Haldane range $-\pi/4<\theta < +\pi/4$.
We have shown that also the dynamical properties do change 
qualitatively and not only quantitatively. Curiously enough the special
point characterizing the change $\theta_{c}$ is distinct from the VBS point.
The bilinear-biquadratic S=1 chain is thus an interesting counterexample
showing the variety of physical phenomena that occur within a massive 
phase.

Finally we have shown evidence that there is no spinon deconfinement when we 
approach the SU(3) phase transition at $\theta =+\pi/4$.
Thus, the Haldane phase is an example of a gapped spin liquid with 
incommensurate spin excitations, a phenomenon that may also appear
in the normal phase of the underdoped cuprates.

\acknowledgments

We thank G. F\'ath and T. Garel for useful discussions, and 
IDRIS (Orsay) for allocation of CPU time on the C94 and C98
CRAY supercomputers and rs6000 workstations.




\end{document}